\documentclass[10pt,conference]{IEEEtran}
\IEEEoverridecommandlockouts
\usepackage{cite}
\usepackage{microtype} %
\usepackage{amsmath,amssymb,amsfonts,amsthm}
\usepackage{algorithmic}
\usepackage{graphicx}
\usepackage{textcomp}
\usepackage{xcolor}
\usepackage{hyperref}
\usepackage{mathtools}
\usepackage{caption}
\usepackage{subcaption}

\newtheorem{example}{Example}
\let\oldexample\example
\renewcommand{\example}{\oldexample\itshape}

\def\BibTeX{{\rm B\kern-.05em{\sc i\kern-.025em b}\kern-.08em
    T\kern-.1667em\lower.7ex\hbox{E}\kern-.125emX}}

\begin{document}
\title{\vspace{-8pt}Improving Figures of Merit for Quantum~Circuit~Compilation}%

\author{
    \IEEEauthorblockN{
        Patrick Hopf\IEEEauthorrefmark{1}, Nils Quetschlich\IEEEauthorrefmark{1}, Laura Schulz\IEEEauthorrefmark{2}, and Robert Wille\IEEEauthorrefmark{1}
    }
    \IEEEauthorblockA{\IEEEauthorrefmark{1}%
        Chair for Design Automation,
        Technical University of Munich,
        Munich, Germany
    }
    \IEEEauthorblockA{\IEEEauthorrefmark{2}%
        QCT Department,
        Leibniz Supercomputing Centre,
        Garching, Germany
    }
    patrick.hopf@tum.de, nils.quetschlich@tum.de, schulz@lrz.de, robert.wille@tum.de\\
    \href{https://www.cda.cit.tum.de/research/quantum}{www.cda.cit.tum.de/research/quantum}
}
\hypersetup{ %
    pdftitle={An Abstract Model and Efficient Routing for Logical Entangling Gates on Zoned Neutral Atom Architectures},
    pdfsubject={IEEE International Conference on Quantum Computing and Engineering},
    pdfauthor={Yannick Stade, Ludwig Schmid, Lukas Burgholzer, Robert Wille}
}

\maketitle

\begin{abstract}

Quantum computing is an emerging technology that has seen significant software and hardware improvements in recent years.
Executing a quantum program requires the compilation of its quantum circuit for a target \emph{Quantum Processing Unit}~(QPU).
Various methods for qubit mapping, gate synthesis, and optimization of quantum circuits have been proposed and implemented in \emph{compilers}.
These compilers try to generate a quantum circuit that leads to the best \emph{execution quality}---a criterium which is usually approximated by \emph{figures of merit} such as the number of (two-qubit) gates, the circuit depth, expected fidelity, or estimated success probability.
However, it is often unclear how well these figures of merit represent the \emph{actual} execution quality on a QPU.

In this work, we investigate the correlation between established figures of merit and actual execution quality on real machines---revealing that the correlation is weaker than anticipated and that more complex figures of merit are not necessarily more accurate.
Motivated by this finding, we propose an improved figure of merit (based on a machine learning approach) that can be used to predict the expected execution quality of a quantum circuit for a chosen QPU without actually executing it.
The employed machine learning model reveals the influence of various circuit features on generating high correlation scores.
The proposed figure of merit demonstrates a strong correlation and outperforms all previous ones in a case study---achieving an average correlation improvement of $49\%$.
\end{abstract}

\section{Introduction}
Quantum computing has made remarkable progress in recent years, with improvements in both the software and hardware used to run programs on quantum computers. 
A quantum program is typically represented as a quantum circuit, composed of a sequence of operations.
For a quantum program to run on a specific \emph{Quantum Processing Unit} (QPU), it needs to be translated into a form that the hardware can execute.
This process is known as quantum circuit \emph{compilation}.

The quality of a compiled quantum circuit is usually measured by so-called \emph{figures of merit}.
Established figures of merit include the number of gates in the circuit, its depth, or the expected fidelity and \emph{Estimated Success Probability} (ESP, ~\cite{esp-lifetime-min}).
These figures of merit are intended to describe how well the circuit will perform on a target QPU. 
However, while these figures of merit provide an approximation of the circuit's \emph{execution quality}, they might not always give an accurate picture of how well the circuit will actually run on quantum hardware. 
QPUs are complex systems that face many challenges during execution, such as interference between signals applied to neighboring qubits, errors during gate or measurement operations, and other hardware imperfections that can impact their performance.
These effects are often difficult to capture with the simple figures of merit used today.

In this paper, we take a closer look at the established figures of merit and investigate how well they truly reflect the quality of a circuit’s execution on real QPUs. 
We find that the correlation between these metrics and real execution performance is often weaker than expected.
In some cases, it even turns out that a more complex metric (like ESP) does not lead to a better approximation. 

Furthermore, to address these weaknesses of established figures of merit, we propose a new way of evaluating circuit quality using machine learning techniques. 
This results in an improved figure of merit that takes into account a variety of quantum circuit characteristics without requiring QPU calibration data.
The approach achieves an average correlation improvement of $49\%$, accurately predicting how well a circuit can be executed on a targeted QPU.
By offering a simple yet more effective method to assess circuit quality, this work helps researchers and engineers to develop or adjust compilers, so that the generated circuits are better suited for a given target QPU.

This paper is structured in the following way:
\autoref{sec:compilation} offers a concise review of quantum circuit compilation including its primary tasks and the established figures of merit for assessing circuit quality. 
\autoref{sec:motivation} discusses the limitations of current metrics (providing the motivation of this work) and presents the proposed approach for an improved evaluation of circuit execution quality. 
\autoref{sec:implement} provides detailed insights into the implementation of the proposed method. 
The results of a study are presented in \autoref{sec:eval}, along with a comparison of the new approach to established figures of merit and a thorough discussion. 
Finally, \autoref{sec:conclusion} summarizes the findings.

\begin{figure*}[htb]
    \centering
    \hfill
    \begin{subfigure}[b]{0.475\textwidth}
        \centering
        \includegraphics[height=55pt]{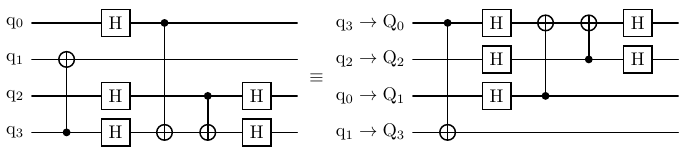}
        \caption{Qubit mapping}
        \label{fig:compilation-a}
    \end{subfigure}
    \hfill
    \begin{subfigure}[b]{0.475\textwidth}
        \centering
        \includegraphics[height=55pt]{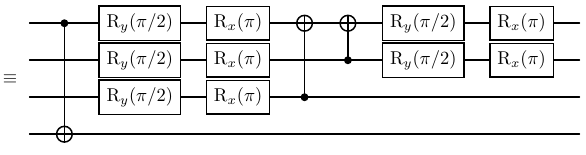}
        \caption{Gate synthesis}
        \label{fig:compilation-b}
    \end{subfigure}
    \hfill

    \hfill
    \begin{subfigure}[b]{0.475\textwidth}
        \centering
        \includegraphics[height=53pt]{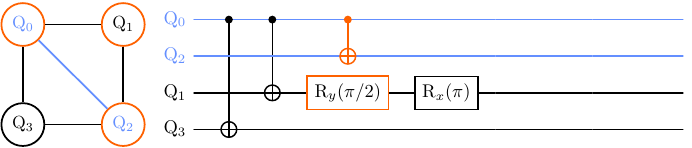}
        \caption{Circuit optimization}
        \label{fig:compilation-c}
    \end{subfigure}
    \hfill
    \begin{subfigure}[b]{0.475\textwidth}
        \centering
        \includegraphics[height=53pt]{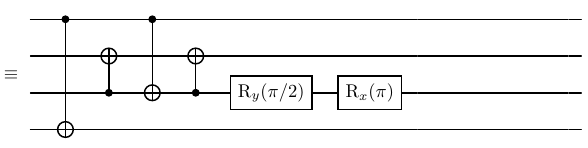}
        \caption{QPU-specific optimization}
        \label{fig:compilation-d}
    \end{subfigure}
    \hfill
    \caption{
        Compilation of a quantum circuit demonstrating (a) mapping, (b) synthesis, and (c), (d) optimization passes for a four-qubit square layout (only missing a link between $\mathrm{Q}_{1}$ and $\mathrm{Q}_{3}$).
        The exemplary QPU is subject to crosstalk errors from parallel gate execution (orange) on neighboring qubits and provides only a low $\mathrm{CNOT}$ fidelity (blue) between distant qubits $\mathrm{Q}_{0}$ and $\mathrm{Q}_{2}$.
    }
    \label{fig:compilation}
\end{figure*}

\section{Quantum Circuit Compilation}\label{sec:compilation}

A quantum program is usually designed as a quantum circuit in order to execute it on quantum hardware.
Such a circuit typically consists of multiple operations called quantum gates and measurements.
During the execution of a program, these operations modify the state of quantum bits, so-called \emph{qubits}---the fundamental computational elements of a QPU.
There are various quantum computing hardware technologies that realize these operations and qubits in different ways (e.g., superconducting, trapped ions, neutral atoms, etc).
Since quantum circuits are typically defined on a hardware-agnostic level, these need to be translated into machine-executable operations.
Any QPU supports a specific set of executable operations. 
Consequently, quantum circuit compilation is necessary to convert any given quantum circuit into an equivalent one that utilizes only these supported operations.
This section provides an overview of the main tasks involved in quantum circuit compilation, along with a review of how the quality of a compiled circuit is currently assessed using so-called figures of merit.

\subsection{Compilation Tasks}

Depending on the type of the QPU and its specific constraints, the compilation procedure usually involves a combination of the following tasks:

\textbf{Qubit mapping}:
Quantum circuits typically contain \mbox{multi-qubit} gates acting on multiple arbitrary qubits.
However, some hardware (like superconducting QPUs) only provides a limited set of qubits on which \mbox{multi-qubit} operations are possible.
Similarly, for technologies (like trapped ions and neutral atoms) that do not have this limitation, it is often sensible to perform them on specific qubits only (e.g., to reduce shuttling operations). 
Executing the algorithm, therefore, usually requires a mapping between the logical (program) qubits and the physical (QPU) qubits.

\begin{example}
    \autoref{fig:compilation-a} shows how the qubits of an example circuit are mapped to an exemplary superconducting QPU architecture with a square qubit layout where almost all physical qubits $\mathrm{Q}_0$, $\mathrm{Q}_1$, $\mathrm{Q}_2$, and $\mathrm{Q}_3$ are connected; only missing a direct link between $\mathrm{Q}_1$ and $\mathrm{Q}_3$.
\end{example}

\textbf{Gate synthesis}:
Quantum algorithms and their corresponding circuits are commonly designed using a wide set of gate and measurement operations.
Since any QPU only supports a small number of natively executable operations, each non-native operation must be synthesized into one or more of these supported operations.
This task is non-trivial, and doing it optimally is NP-complete~\cite{pehamDepthoptimalSynthesisClifford2023}.

\begin{example}
    \autoref{fig:compilation-b} illustrates how the Hadamard gates in the circuit of the previous example can be translated into the native $R_x$ and $R_y$ rotation gates provided by the exemplary superconducting QPU architecture.
\end{example} 

\textbf{Circuit optimization}:
It is possible to alter a circuit's gate composition without changing its original function.
Following specific transformation rules, a general quantum circuit can be expressed through numerous combinations of distinct gates.
Hence, a circuit can often be optimized with respect to a desired metric.

\begin{example}
    \autoref{fig:compilation-c} demonstrates how the four \mbox{single-qubit} rotation gates of the previously unoptimized circuit can be eliminated (according to circuit transformation rules~\cite{equivalent-circs}).
\end{example}

Various compilation methods have been proposed for qubit mapping~\cite{
    lin_layout_synthesis, zulehnerEfficientMethodologyMapping2019, matsuoEfficientMethodQuantum2019, bochen_mapping, willeMappingQuantumCircuits2019, liTacklingQubitMapping2019, pehamOptimalSubarchitecturesQuantum2023, hillmichExploitingQuantumTeleportation2021, bqskit_mapping, tket_mapping, schmidHybridCircuitMapping2024, MQT_QMAP, Paraskevopoulos2023
}, gate synthesis~\cite{
    gilesExactSynthesisMultiqubit2013, amyMeetinthemiddleAlgorithmFast2013, millerElementaryQuantumGate2011, zulehnerOnepassDesignReversible2018, niemann2020advancedexactsynt, pehamDepthoptimalSynthesisClifford2023, bqskit_opt_syn
}, and circuit optimization~\cite{
    bqskit_opt_syn, hattoriQuantumCircuitOptimization2018, vidalUniversalQuantumCircuit2004, itokoQuantumCircuitCompilers2019, maslovQuantumCircuitSimplification2008, niu2023powerfulquantumcircuitresizing, Staudacher_2023,sünkel2023ga4qcogeneticalgorithmquantum,ibm_rl_synthesis
}.
These tasks are usually implemented by individual compilation passes that manipulate the circuit.
Passes can be performed in any order and might be repeated multiple times.
Hence, there are various pass sequences that lead to distinct compiled versions representing the same original circuit.
Finding a suitable order of compilation passes that yield an efficient circuit is a non-trivial task.
This raises the question: How can the quality of a chosen sequence of passes and its associated quantum circuit be assessed?

\subsection{Figures of Merit}\label{sec:FoM}

Besides imposing hardware-specific constraints, current QPUs additionally pose the risk of erroneous calculations.
Quantum hardware is usually subject to environmental noise and suffers from imperfect gate and qubit realizations.
Due to this, any quantum operation (even the identity that should leave a qubit unmodified) can only be performed with some probability of error.
To obtain a circuit that produces high-quality execution results, its design must minimize the accumulation of errors.
Figures of merit can act as a proxy for the result quality, enabling the assessment of a compilation run without the need to execute the resulting circuit.
Thus far, the following figures of merit have been employed in \mbox{state-of-the-art} mapping, synthesis, and optimization methods such as those mentioned above.

\begin{itemize}
    \item \textbf{Number of gates}, i.e.,
    the integer gate count in the circuit.
    Often, only two-qubit gates are considered because of their dominant error rates.
    A lower number indicates better performance.

    \item \textbf{Circuit depth}, i.e.,
    the integer number of gates on the longest path in the circuit graph.
    A lower circuit depth usually indicates lower execution time and fewer gates, hence a higher circuit quality.

    \item \textbf{Expected fidelity}, i.e.,
    the product of all decimal gate and measurement fidelities in the circuit.
    Since the fidelity is inversely proportional to the error, higher values suggest better execution quality.
    \item \textbf{Estimated Success Probability} (ESP), i.e., the expected fidelity multiplied by the exponential decay factor
    \(
        \exp{\left[- t^q_\mathrm{idle}/\min{(T^q_1, T^q_2)} \right]}, 
    \)
    for all qubits, where $t^q_\mathrm{idle}$ is the total idle time of qubit $q$.
    This figure additionally requires each qubit's $T_1$ and $T_2$ relaxation times (which measure how long it can retain information) and is based on variants of ESP~\cite{esp-lifetime-min, esp-idle, Schmid_2024}.
    High values indicate good performance.
\end{itemize}

The first two figures of merit are hardware-agnostic metrics and, hence, independent of the executing QPU.
The latter two require experimental data about the specific hardware---usually obtained during device calibration, a process that involves \mbox{fine-tuning} qubits, gates, and measurement fidelities.
The intuition behind using these as proxies for a circuit’s execution quality is that these numbers are expected to scale (directly or indirectly) proportional to the anticipated errors.
Under this assumption, most quantum compilation flows optimize for one (sometimes multiple) of these figures of merit.
However, it often remains unclear whether the resulting compiled circuit actually leads to the least error-affected execution and, therefore, the best solution.

\section{Motivation and Proposed Approach}\label{sec:motivation}

The compilation concepts reviewed before provide an \mbox{easy-to-use} and general approach to converting any algorithm encoded as a quantum circuit into an executable set of operations.
However, the simplicity and generality can come at the cost of missing out on better circuit designs, as demonstrated in the following example.

\begin{example}\label{sample}
    The previously considered circuit, depicted in \autoref{fig:compilation-c}, is compiled according to the established figures of merit reviewed before, i.e., is minimized with respect to the overall number of gates and, accordingly, the circuit depth.
    Depending on the actual gate fidelity and relaxation values, this solution will also maximize the expected fidelity and ESP.

    However, considering the QPU in \autoref{fig:compilation}(c), this circuit is unnecessarily prone to crosstalk on neighboring qubits $\mathrm{Q}_{0}$, $\mathrm{Q}_{1}$, and $\mathrm{Q}_{2}$---an error that occurs when gates are executed in parallel (highlighted in orange).
    This effect, along with a low CNOT fidelity between distant qubits $\mathrm{Q}_{0}$ and $\mathrm{Q}_{2}$ (highlighted in blue), can be avoided by using the functionally equivalent circuit shown in \autoref{fig:compilation}(d), where two CNOTs are rearranged and an additional one is added.
    This circuit has a higher number of gates and depth (and would, therefore, be rejected when applying established figures of merit) but still performs better when executed on the considered QPU. 
    While expected fidelity and ESP can account for the low CNOT fidelity between distant qubits, they remain indifferent to the crosstalk effects.
\end{example}

The example illustrates how relying on the established figures of merit, which may not fully capture hardware-specific characteristics, can guide the compilation procedure to subpar solutions.
Similar concerns have been raised before~\cite{lubinski2024quantumalgorithmexplorationusing, Venturelli2019QuantumCC}, and were confirmed in a study demonstrating that calibration-based compilation strategies can achieve higher circuit fidelities compared to those that solely focus on minimizing the number of two-qubit gates \cite{kurniawan2024usecalibrationdataerroraware}.
Likewise, an individual assessment of ESP demonstrated its poor correlation with actual device performance~\cite{esp-lifetime}.
This already indicates the need for a comprehensive investigation (and, eventually, improved figures of merit), but to the best of our knowledge, no comprehensive study that directly compares the correlation scores of various established figures of merit and circuit execution quality has been conducted yet.

At the same time, the investigation and development of alternative figures of merit are still in the early stages.
New figures of merit have been introduced employing basic machine learning techniques \cite{optimizability, Vadali_2023}, where the underlying circuit representations scale with the depth of the input \mbox{circuit---making} these methods impractical for deep quantum circuits.
This issue is also present in another approach utilizing the circuit graph representation, which was used by a \mbox{transformer-based} model to accurately predict the probability of successful trials \cite{quest}.
Although more sophisticated, this work only considered circuits of up to seven qubits and (like ESP) requires accurate $T_1$, $T_2$, gate and measurement fidelity data, which is often outdated or not available.

In summary, there is a lack of comprehensive analysis of the established figures of merit, while emerging alternatives struggle with scalability and practical limitations.
In this work, we address these gaps with the following contributions:
\begin{enumerate}
    \item 
    We conduct a comprehensive investigation to quantify the (weak) correlation between the established figures of merit (i.e., number of gates, circuit depth, expected fidelity, and ESP) and a circuit's actual execution quality.
    The study is designed with a focus on real-world applicability by executing circuits from practical quantum computing applications on real QPUs.
    \item 
    Based on these findings, we propose an interpretable machine-learning-based figure of merit as an improved representation of a circuit's execution quality.
    This model works with a depth-independent circuit representation and provides an individualized figure of merit for any QPU without requiring detailed calibration data.
\end{enumerate}

\section{Implementation}\label{sec:implement}

This section provides details on the implementation of the contributions outlined above.
First, we introduce the measure required to evaluate the execution quality of a quantum circuit and demonstrate its correlation with the previously introduced figures of merit.
Based on this measure, a machine learning approach to generate an improved figure of merit is proposed, which offers a better correlation and, thus, provides a more accurate approximation of the execution quality for a given quantum circuit.

\subsection{Investigating the Correlation Between \newline Figures of Merit and Execution Quality}

\begin{figure}[t]
\centering
\includegraphics[width=\linewidth]{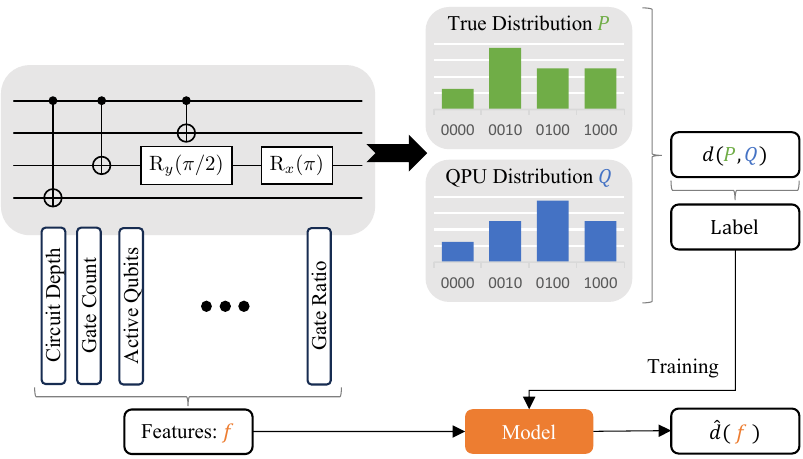}
\caption{Workflow for feature and label generation from a compiled quantum circuit. The Hellinger distance---representing the difference between the circuit's true distribution and the QPU execution results---is used as label data for model training.}
\label{fig:setup}
\vspace{-6pt}
\end{figure}

The measurement result of a quantum circuit is usually described in terms of a discrete probability distribution over all possible qubit states, i.e., combinations of zeroes and ones that can be illustrated in a histogram (see green and blue charts in \autoref{fig:setup}).
In order to understand how well a figure of merit represents the presumed execution quality of a quantum circuit, we evaluate the result quality of its execution on an actual QPU and compare it to its \emph{true distribution}.
The true (noiseless) distribution can be obtained, e.g., from a state vector simulation, whereas the (noisy) experimental distribution can be obtained from repeated circuit executions on a QPU. 
To quantify the execution quality and, accordingly, the (mis)alignment of the two histograms, the \emph{Hellinger distance}
\begin{equation}\label{Hellinger}
    d(P, Q) = \frac{1}{\sqrt{2}} \sqrt{\textstyle \sum_{i=0}^{2^N-1} \left(\sqrt{p_{|i\rangle}} - \sqrt{q_{|i\rangle}} \right)^2} \in [0, 1]
\end{equation}
between the true distribution $P = \{p_{|0\rangle}, \dots, p_{|2^N-1\rangle} \}$ and its QPU counterpart $Q = \{ q_{|0\rangle}, \dots, q_{|2^N-1\rangle} \}$ is used.
If the measurement histograms of the true and experimental QPU distribution overlap completely, their distance is zero. 
Conversely, for highly distinct histograms, the distance approaches one.

In addition to assessing the Hellinger distance $d$, we investigate its correlation with any previously introduced figure of merit $y$ on a set of $M$ quantum circuits.
For this task, the \emph{Pearson correlation} coefficient
\begin{equation}\label{Pearson}
    r = \frac{\sum_{j=1}^M (d_j - m_d)(y_j - m_y)}{\sqrt{\sum_{j=1}^M (d_j - m_d)^2 \sum_{j=1}^M (y_j - m_y)^2}} \in [-1, 1]
\end{equation}
is calculated, where $m_d$ and $m_y$ are the mean (Hellinger distance $d$ and figure of merit $y$) values over all circuits in the set. 
A perfect linear correlation is represented by $|r| = 1$, whereas $r = 0$ indicates no Pearson correlation at all\footnote{%
There might be non-linear correlation measures that better capture the relationship between individual figures of merit and the Hellinger distance. 
However, any such measure must include a linear (or anti-linear) component that the Pearson correlation coefficient can capture.
}.

This correlation (based on the Hellinger distance) can now be used to quantify how well any figure of merit actually approximates the execution quality.
Furthermore, the Hellinger distance is additionally used to derive an improved (machine-learning-based) figure of merit that aims to capture it more accurately and, thus, can be used as a more precise figure of merit.

\subsection{Proposed Figure of Merit}

With insights from the Hellinger distance, it is possible to quantify the (mis)alignment between the results obtained from executing a circuit on a real QPU and the true distribution.
While established figures of merit use indirect metrics to approximate this measure in order to guide the circuit compilation, it would be far more efficient to directly optimize for a reduction of the Hellinger distance.
However, evaluating the distance for every possible circuit configuration during compilation would require an impractical amount of simulation and execution data.

Hence, we instead propose to train a machine learning model on a representative set of practical algorithms labeled with experimentally obtained Hellinger distance values.
The resulting model then acts as an estimator to predict the distance for any circuit during compilation, effectively serving as a figure of merit.

To this end, we employ the workflow depicted in \autoref{fig:setup}.
In order to train an estimator model (pictured in orange) for a specific QPU, a comprehensive set of feature and label data is required.
Such an estimator receives as input a vectorized representation, called \emph{feature vector} (shown in the bottom left), of all the quantum circuits.
To this end, we utilize a revised version of the circuit encoding introduced in~\cite{quetschlich2024mqtpredictor}, whose size is independent of the circuit depth and, therefore, constant for any specific QPU.
Among the basic features are the hardware-agnostic figures of merit, i.e., the circuit depth and its gate counts.
More sophisticated features include circuit liveness, which captures how actively qubits are utilized; directed program communication, which quantifies the ratio between the actual and maximum possible average node degree of the circuit's directed interaction graph; as well as parallelism~(all based on\cite{supermarq}) and gate ratios, which reflect the circuit's operational density.
Notably, the feature vector does not require calibration data, as is required for calculating fidelity-related figures of merit.

In addition to extracting the feature representation, every single circuit must be associated with its Hellinger distance, which serves as the training label (shown in the top right).
This requires evaluating the noiseless true result distribution and the noisy QPU distribution, as shown in the green and blue sample Histogram.

Given a representative set of such circuit features and label data, a model can be trained and then used to estimate the Hellinger distance for a given circuit, thereby allowing the compilation to  aim directly at the reduction of Hellinger distance.

\section{Experimental Evaluation}\label{sec:eval}
The ideas and implementation details described above eventually lead to a framework that allows for (1) a comprehensive investigation of the correlation between circuit execution quality and the established figures of merit and (2) an evaluation of the proposed (improved) figure of merit.
In this section, we summarize the main results obtained by these investigations and evaluations.
To this end, we first review the used setup.
Afterwards, the obtained results are provided and discussed.

\subsection{Setup}
The following describes the experimental setup used in our investigations and evaluations.
All of the presented methods and results were implemented in Python and are accessible through the MQT Predictor~\cite{quetschlich2024mqtpredictor} as part of the \emph{Munich Quantum Toolkit}~\cite{mqt}. 
The source code is publicly available on GitHub\footnote{\vspace{-5pt}
    https://github.com/cda-tum/mqt-predictor
}.

\subsubsection{Used Benchmarks}
For the comprehensive investigation of quantum circuit quality, we utilize all circuits provided by the MQT~Bench collection, an open-source library frequently used to evaluate compilers, QPUs, and more~\cite{mqt-bench}. 
This collection offers a variety of algorithms (like VQE, QAOA, QFT, etc.), which have been mapped, synthesized, and optimized for any number between $2$ and $20$ qubits using the Qiskit~\cite{qiskit} transpiler module at optimization level three.
Since circuits with a depth of more than $1000$ are too much affected by noise when executed on current quantum computers (and, eventually, would not produce any meaningful results), we only considered circuits with a compiled depth smaller than $1000$---leaving a total of $222$ circuits. 

\subsubsection{Used QPUs}
The resulting set of benchmark circuits has been executed on two superconducting IQM QPUs hosted at the German Leibniz Supercomputing Centre.
Both devices are members of the $20$-qubit series (labeled Q20-A and Q20-B)~\cite{iqm}.
Their native gate set consists of a parameterized single-qubit rotation gate and the CZ gate on IQM's crystal architecture (qubits located on a square grid).
In addition to running them on both QPUs (and generating the corresponding true distributions), the full set of benchmark circuits has been simulated using the Qiskit Aear noiseless state vector simulator on a MacBook Pro (M2 chip), completing within a few hours.

\subsubsection{Machine Learning Model}
All circuits have been expressed through the numeric feature vector of size $30$ and have been labeled with their associated Hellinger distance values.
Then, a random forest regressor (consisting of multiple decision trees, implemented with scikit-learn~\cite{scikit-learn}) was trained for each QPU on the same classical hardware in a few seconds.
This was done using cross-validation over three training sets and an overall \mbox{$80/20$~train-test} ratio. 
The Pearson correlation coefficient served as the model performance score during validation.
A hyperparameter grid search to optimize, e.g., the number of decision trees, their maximum depth, and the minimum samples per leaf and split, could be completed in under a minute on the same classical hardware.
Eventually, like any other figure of merit, the trained model was used to determine the quality of a compiled circuit and evaluated on the (previously unseen) test set.

\subsection{Investigation of Established Figures of Merit}\label{eval:A}

\begin{table}[t]
\caption{Pearson correlation with Hellinger distance}
\begin{center}
\begin{tabular}{l|c|c|c}
    \textbf{Figure of merit / QPU}  & Q20-A           & Q20-B           & Combined          \\
    \hline &&& \\[-7pt]
    Number of gates                 & $0.46$          & $0.61$          & $0.53$            \\
    \hline &&& \\[-7pt]
    Circuit depth                   & $0.46$          & $0.62$          & $0.54$            \\
    \hline &&& \\[-7pt]
    Expected fidelity               & $0.66$          & $0.80$          & $0.73$            \\ 
    \hline &&& \\[-7pt]
    ESP                             & $0.59$          & $0.70$          & $0.64$            \\
    \hline &&& \\[-7pt]
    \textbf{Proposed approach}      & $\mathbf{0.88}$ & $\mathbf{0.94}$ & $\mathbf{0.91}$   \\
\end{tabular}
\label{tab}
\end{center}
\vspace{-16pt}
\end{table}

After executing the entire benchmark set on both QPUs and generating the true distributions, the correlation between the established figures of merit and the actual circuit execution quality has been evaluated. 
The results are summarized in \autoref{tab}, showing the Pearson coefficients for each investigated figure of merit.
Values in the columns \emph{Q20-A} and \emph{Q20-B} correspond to the executing QPUs, whereas the values in the column \emph{Combined} provide the correlation for all circuit executions on both QPUs.
To enhance clarity, the table only shows the absolute correlation scores.
Values closer to $1$ indicate higher quality figures of merit.

The results provide some interesting insights (both expected as well as unexpected): 
First, they show that the number of gates and circuit depth have very similar correlation scores, which, considering their clear link between each other, is not really surprising. 
Furthermore, the expected fidelity and ESP obviously provide significantly higher correlation values than the other figures on both QPUs. 
Also, this is not surprising: The number of gates and the circuit depth are rather simple (albeit easy to use) figures of merit, while expected fidelity and ESP take hardware information into account. 
Hence, a better quality is expected from these figures of merits.
 
What surprises, though, is that, in some cases, a more complex metric does not necessarily lead to a better correlation.
In fact, even though expected fidelity and ESP share the same fidelity term in their calculation, the former achieves a higher, i.e., better correlation score ($0.66$ vs. $0.59$ on \mbox{Q20-A} and $0.80$ vs. $0.78$ on \mbox{Q20-B}). 
Since the only difference lies within the calibration-data-dependent relaxation term, this result points to possibly outdated $T_1$, $T_2$ times.
 
Independently from those differences, the results confirm that \emph{all} established figures of merits \emph{do not} provide a fully effective correlation between estimated and real execution performance.
Even though the hardware-specific figures of merit (i.e., expected fidelity and ESP) represent the actual circuit execution quality better than the target-agnostic figures of merit (i.e., number of gates and circuit depth), the combined correlation remains at $0.73$ in the best case.
This confirms the discussions and the motivation from~\autoref{sec:motivation}, highlighting that the established figures of merit indeed leave room for improvement.

\subsection{Evaluation of Proposed Figure of Merit}\label{eval:B}
The above investigation confirmed the weaknesses of the established figures of merit.
Next, we evaluated whether the proposed figure of merit provides an improvement.
To this end, the correlation of the trained machine learning model is assessed using the unseen circuit test set.
Its Pearson correlation is presented in the last row of \autoref{tab}, again for both individual QPUs and for the total set of all executed test circuits.

The results clearly confirm the improved correlation of the proposed figure of merit to the actual execution quality.
In fact, on average, the correlation score increases by $62\%$ and $38\%$ for the Q20-A and Q20-B devices, respectively. 
Considering the average correlation of all previous figures of merit over both QPUs (last column), the proposed figure of merit outperforms their correlation scores by $49\%$.

\begin{figure}[t]
\centering
\vspace{-5pt}
\includegraphics[width=\linewidth]{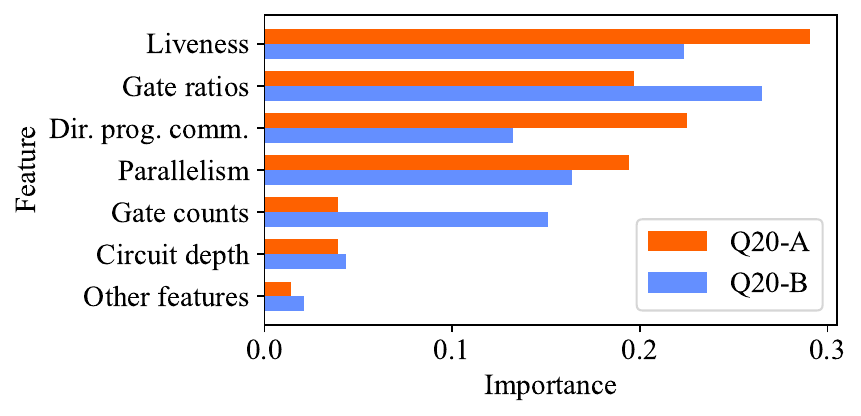}
\caption{Random forest model feature importance.}
\label{importance}
\vspace{-15pt}
\end{figure}

In order to understand how the proposed figure of merit managed to capture the execution quality so well, we investigate the model's feature importance depicted in \autoref{importance}.
It can be observed that the model's prediction quality strongly depends on features designed to capture qubit activity, operational density, and qubit interactions---specifically, liveness, gate ratios, parallelism, and directed program communication.
In contrast, the basic gate counts and circuit depth features show moderately low importance, aligning with the correlation values observed in the previous figure of merit analysis.

Overall, the proposed figure of merit offers a significant improvement over the established figures of merit.
By leveraging the right combination of circuit features, it manages to capture the actual circuit execution quality much better. 

\subsection{Discussion}
The findings presented above provide valuable insights into the characteristics of an effective figure of merit, enhancing our understanding of both hardware-agnostic and hardware-specific (calibration-data-based) approaches.
The discrepancy between expected fidelity and ESP indicates that overly specific metrics can reduce the correlation when using poor calibration data.
This result is consistent with work on error-aware compilation methods, which has also found decreased compilation performance on outdated calibration data \cite{kurniawan2024usecalibrationdataerroraware}. 

Furthermore, the results showed that neither the number of gates nor the circuit depth alone serves as a reliable estimator of circuit execution quality.
This finding is supported by their relatively low contribution to the random forest model accuracy.
Accounting for individual qubit performance through idle times in the exponential decay factor (see ESP) did not improve correlation scores. 
However, incorporating their impact via liveness and parallelism features did.  
This underscores the importance of selecting and combining circuit features effectively, aligning with similar findings in related work \cite{bandic2024profilingquantumcircuitsefficient, Bandic_2023}, and demonstrates that such a feature set can yield a far better figure of merit than any single measure alone.

Finally, unlike the expected fidelity and ESP, the proposed model does not directly rely on \mbox{device-specific} measurement data, which is highly valuable when this information is not (frequently) provided by a QPU provider.
Importantly, the model was trained on real QPU data rather than simulations, which means it indirectly incorporates device-specific calibration information. 

Overall, these findings show that it is crucial to find the right balance between incorporating an accurate hardware representation and abstracting device details.
The high correlation scores obtained through the proposed figure of merit indicate that the model managed to achieve this performance requirement.
Future work will focus on examining the model's performance over time, comparing it to other QPU-specific figures of merit in the context of evolving QPU noise characteristics.

Lastly, in our experiments, we trained the model on circuits that can still be classically simulated. 
With improving hardware, this will become more challenging. 
However, there is evidence to suggest that the \emph{Probability of Successful Trials}~(PST) derived by appending a circuit's inverse (hence, removing the need for simulation) can successfully represent its execution quality~\cite{quest}.
Future work will investigate to what extent the PST can be used to improve our proposed approach.

\section{Conclusion}\label{sec:conclusion}

This work investigated and improved upon the limitations of current figures of merit in representing the actual execution quality of quantum circuits. 
By analyzing the correlation between established figures of merit---such as the number of gates, circuit depth, expected fidelity, as well as ESP---and real-world execution results, we unveiled that these figures of merit often fall short of accurately representing quantum circuit performance on a QPU.
This gap highlights the need for improved metrics that better align with execution quality.

Motivated by that, we introduced a machine-learning-based figure of merit designed to better correlate with actual circuit execution quality. 
The proposed model does work with a \mbox{depth-independent} circuit representation and provides a \mbox{QPU-specific} figure of merit. 
The method outperformed the traditional figures of merit, showing a $49\%$ improvement in its correlation with execution quality.

Overall, the obtained findings underscore the significance of selecting and combining the right circuit characteristics to develop a figure of merit that closely aligns with actual circuit execution quality. 
This work demonstrates that the appropriate format and combination of circuit features can yield a far superior figure of merit than any individual measure alone. 
We have shown that, to this end, machine learning can significantly enhance quantum circuit compilation, providing a more effective approach to evaluating execution quality.

\section{Acknowledgements}
P.H., N.Q., and R.W. acknowledge funding from the European Research Council (ERC) under the European Union’s Horizon 2020 research and innovation program (grant agreement No. 101001318), the Munich Quantum Valley, which is supported by the Bavarian state government with funds from the Hightech Agenda Bayern Plus, and has been supported by the BMWK on the basis of a decision by the German Bundestag through project QuaST, as well as by the BMK, BMDW, the State of Upper Austria in the frame of the COMET program, and the QuantumReady project within Quantum Austria (managed by the FFG).
L.S. acknowledges funding by the German Federal Ministry for Education and Research under grants 13N15689 (DAQC), and 13N16063 (Q-Exa).
\vspace{200mm}

\bibliographystyle{ieeetr}
\bibliography{references}
\end{document}